\documentstyle[epsfig,aps,prb,floats,twocolumn]{revtex}
\begin{document}

\title{Dynamics of lattice pinned charge stripes}
\author{Yu.\ A.\ Dimashko$^a$, C.\ Morais Smith$^{a,b}$, 
N.\ Hasselmann$^{a,c}$, and A.\ O.\ Caldeira$^d$}
\address{$^{a \,}$I Institut f{\"u}r Theoretische Physik, Universit{\"a}t 
Hamburg, D-20355 Hamburg, Germany \\
$^{b \,}$Institut de Physique Th{\'e}orique, Universit{\'e} de Fribourg, 
P{\'e}rolles, CH-1700 Fribourg, Switzerland \\
$^{c \,}$Dept. of Physics, University of California, Riverside, CA, 92521, 
USA \\
$^{d \,}$Instituto de F{\'\i}sica Gleb Wataghin, Universidade Estadual de 
Campinas, CP 6165, 13085-970 Campinas SP, Brasil \\
\rm{(\today)}\bigskip\\
\parbox{14.4cm}{\rm
We study the transversal dynamics of a charged stripe (quantum string)
and show that zero temperature quantum fluctuations are able to depin it
from the lattice.  
If the hopping amplitude $t$ is much smaller 
than the string tension $J$, the string is pinned by the underlying lattice. 
At $t \gg J$, the string is depinned and allowed to move freely, if we 
neglect the effect of impurities. 
By mapping the system onto a 1D array of Josephson junctions, we show that 
the quantum depinning occurs at $(t/J)_c = 2 / \pi^2$. Besides, we exploit 
the relation of the stripe Hamiltonian to the sine-Gordon (SG) theory and 
calculate the infrared excitation spectrum of the quantum string for 
arbitrary $t/J$ values.
 }}

\maketitle
\narrowtext

The existence of a striped phase in doped 2D antiferromagnets (AF) has been
recently a subject of intense experimental and theoretical investigations.
Experimentally, elastic \cite{Woch,Tran} and inelastic \cite{Yama} neutron
diffraction measurements in nickelates \cite{Woch} and cuprates \cite
{Tran,Yama} have revealed the presence of charge and spin-order. Besides,
muon spin resonance and nuclear quadrupole resonance results \cite{Bors}
have also been successfully interpreted within the picture of charged domain
walls separating antiferromagnetic domains. Striped phases have repeatedly
been found in numerical investigations of $t-J$ and Hubbard models.\cite{tJH} 
It is possible that the striped phase is responsible for
many of the unusual properties of the cuprate superconductors.\cite{Emer}

In the present paper, we study within a phenomenological model 
\cite{Eske,Mora,Hass} the transversal dynamics of a single stripe 
(quantum string). By performing a canonical transformation in the 
quantum string Hamiltonian, we map the system onto a 1D array of Josephson 
junctions, which is known to exhibit an insulator/superconductor 
transition at $(t/J)_c =2 / \pi^2$. This transition is also known to represent
the unbinding of vortex/antivortex pairs \cite{KT} in the
equivalent XY model. Further, by exploiting the relation of these models to
the sine-Gordon (SG) theory \cite{Itzy}, we study the spectrum of the quantum
string in a sector of zero topological charge of its Hilbert space
and reveal the meaning of the transition in the 
``string'' language. At $(t/J)_c$  the (insulating) pinned phase, 
corresponding to an energy spectrum with a finite gap, 
turns into a (metallic) depinned phase where the spectrum becomes gapless.
In doing so, we have connected two important and different classes of 
problems, i.e., the transversal dynamics of stripes in doped AF and a system 
with the well known properties of the SG theory. 

Let us consider a single vertical string on a $N\times L$ square lattice  
(see Fig.\ 1a). The linear concentration of holes in the string is assumed to 
be one hole/site. The string is composed of $N$ charged particles elastically 
interacting with the neighbour ones and constrained to move along $N$ 
horizontal lines. The lattice constant is taken as the unit of length. 

The classical state of the system is described by the  $N$-dimensional vector 
 ${\vec x}= \{x_1,x_2,...,x_N\}$. 
Here, $x_n$ is the $x$-coordinate of the $n$-th particle, $x_n=1,2,...,L$. 
The corresponding quantum state $\mid \vec x >$ is 
defined as an eigenstate of all the coordinate operators  
 $\hat{x}_n$, $n=1,2,...,N$ : $\hat{x}_n \mid \vec{x} > = x_n \mid \vec{x}>$.
The phenomenological Hamiltonian describing this system is 
\begin{equation}
\hat{H}=-t\sum_n \left(\hat{\tau}_n^{+} + \hat{\tau}_n^- \right) + 
\frac{J}{2}\sum_n \left( \hat{x}_{n+1} - \hat{x}_n \right)^2.
\label{Ham1}
\end{equation}

\begin{figure}[t]
\unitlength1cm
\vspace{0.1cm}
\begin{picture}(7,5)
\epsfxsize=5cm
\put(1.5,0.5){\epsfbox{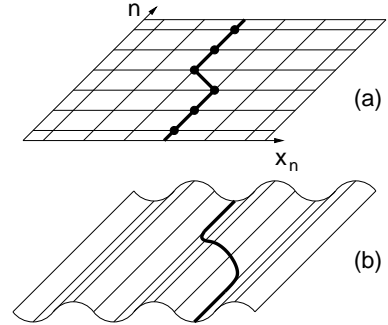}}
\end{picture}
\caption[]{\label{fig1} Two models of stripe: a) discrete string model; 
b) continous sine-Gordon model.}
\end{figure}

The translation operators $\hat{\tau}_n^{\pm}$ are defined
by their action on the coordinate states, 
$\hat{\tau}_n^{\pm} \mid \vec{x} > = \mid \vec x \pm \vec{e}_n >$, 
where  $(\vec{e}_n)_m = \delta _{nm}$. 
The coefficients $t$ and $J$ denote the hopping amplitude and the string 
tension, respectively. The operators $\hat{\tau}_n^{\pm}$ can be expressed 
through the momentum operators $\hat{p}_n$, which obey the canonical relation 
 $[\hat{x}_n,\hat{p}_m ] = i \delta_{nm}$.
We then find $\hat{\tau}_n^{\pm}=\exp (\pm i \hat{p}_n )$ and 
the Hamiltonian (\ref{Ham1}) becomes 
\begin{equation}
\hat{H}=-2t\sum_n \cos \hat{p}_n + \frac{J}{2}
\sum_n \left(\hat{x}_{n+1}-\hat{x}_n \right)^2.
\label{Ham2}
\end{equation}

Hereafter, we classify the state of the quantum string by the value of the 
topological charge $\hat{Q} = \sum_n ( \hat{x}_{n+1}- \hat{x}_n).$ In the case
of open boundary conditions (BC), the topological charge is an arbitrary 
integer, $Q=0,\pm 1,\pm 2,...$ The states with positive and negative charges 
are called kinks (K) and antikinks (AK), respectively. Here, we consider 
periodic BC, $\hat{x}_{N+1}= \hat{x}_1$. Hence, 
the total topological charge of the string is zero.

Since we are interested in the conducting properties of the system, we have
to determine the current operator $\hat{\j}_n = e \dot{\hat{x}}_n$,
where $e$ is the charge of the particle and the dot denotes the time 
derivative. 
Using the equation of motion $\dot{\hat{x}}_n = i[\hat{H}, \hat{x}_n]$, 
we obtain $\hat{\j}_n = 2 e t \sin \hat{p}_n$.

At this point, it is convenient to perform a dual transformation to new
variables referring to the segments of the string, i.e., to a pair of
neighbour holes, 
\begin{equation}
\hat{x}_n - \hat{x}_{n-1} = \hat{\pi}_n, \qquad 
\hat{p}_n = \hat{\varphi}_{n+1} - \hat{\varphi }_n.
\label{dt}
\end{equation}
The new local variables also obey the canonical relation $[\hat{\varphi}_n, 
\hat{\pi}_m ] = i \delta_{nm}$. Furthermore, we take the limit $L \to
\infty$ in order to deal with all operators  in the $\varphi $-representation,
 $ \hat{\varphi }_n \Rightarrow \varphi _n,\, \hat{\pi}_n \Rightarrow -i
\partial /\partial \varphi _n$.
The continuous variable $\varphi_n$ is restricted to the interval
 $0 \leq \varphi_n < 2 \pi$. Finally, the Hamiltonian and the transverse
current operator acquire the form
\begin{eqnarray}\nonumber
\hat{H}=-2t\sum_n\cos (\varphi_{n+1}-\varphi_n) &-& \frac{J}{2}
\sum_n(\partial /\partial \varphi_n)^2, \\
\hat{\j}_n=2et\sin (\varphi_{n+1} &-& \varphi_n),
\label{seg}
\end{eqnarray}
which is known from the theory of superconducting chains.
Eqs.\ (\ref{seg}) describe a Josephson junction chain, with the
Coulomb interaction taken into account. The solution of this problem
at $T = 0$ has been found by Bradley and Doniach \cite{Brad}. Depending
on the ratio $t/J$, the  chain is either insulating (small $t/J$) or
superconducting (large $t/J$). The results arise from the standard mapping
of the 1D quantum problem onto the 2D classical one. One obtains
the XY model with Euclidean action \cite{Itzy}
\begin{equation}
{\cal S}_E = \sqrt{\frac{2t}{J}} \sum_{<\vec{r}, \vec{r}'>} \cos 
\left( \varphi_{\vec{r}} - \varphi_{\vec{r}'} \right),
\label{Ecaction}
\end{equation}
where the vectors $\vec{r} = (n, \tau)$ form a rectangular lattice 
in space and imaginary time.

At $t/J = 2/ \pi^2$ the Josephson chain undergoes a Kosterlitz-Thouless (KT) 
transition. \cite{KT}  For
small $t/J$  values, the two-points correlator 
 $<\exp  i ( \varphi_{\vec r} - \varphi_{\vec r'})  >$ decays exponentially.
Then, the frequency dependent conductivity exhibits a resonance, 
Re $\sigma(\omega) \propto \delta (\omega - J)$. 
Since there is no conductivity at $\omega = 0$, this is an insulating state
with a gap $\Delta = J$. In the opposite case, when $t/J$ is large, the same
correlator decays algebraically. Then, the conductivity is
singular at $\omega = 0$, Re $\sigma (\omega) = 2 \pi e^2 t \delta
(\omega)$, and the array is superconducting. 

These results are also valid for the quantum string on the lattice.
Now, it remains to reveal their physical significance for the striped phase. 
In order to achive this aim, we first analyze the problem in two limiting 
cases: $t\ll J$ and $t\gg J$.

In the limit of weak fluctuations, $t\ll J$, the energy spectrum is discrete 
with spacing $\approx J$. The first excitation is separated from the ground 
level by a gap $\Delta \approx J$. This is the minimal energy required to 
create the doublet excitation K-AK, i.e., to change the initially 
flat configuration of the string. Hence, the ground state is insulating and 
the elementary excitations are pairs of bound K/AK. The dimension of 
the pair can be estimated as the correlation length $\xi =2/\ln (J/2t)\ll 1$. 

In the limit of strong fluctuations, $t\gg J$, we can expand the $\cos$-term 
in the Hamiltonian (\ref{seg}) up to second order $\cos (\varphi_{n+1} - 
\varphi_n) \approx 1 - (\varphi_{n+1} - \varphi_n)^2 /2$ and diagonalize the 
quadratic Hamiltonian. Then, we obtain the phonon-like spectrum 
$E_k=-2tN+\sqrt{8tJ}\mid \sin (k/2)\mid$
with a finite band width $\sqrt{8tJ}$ and no gap. Therefore, the ground state
is conducting and the stripe is depinned. The calculations of the conductivity
are straightforward, since in this case the time dependence of the current  
$\hat{\j}_n \approx 2et (\hat{\varphi}_{n+1} - \hat{\varphi}_{n})$ 
follows from the standard relation, 
\begin{equation} 
\hat{\varphi}_n (\tau) 
= \sum_k \sqrt{\frac{J}{2N\omega_k}}
[e^{i(kn-\omega_k \tau)}\hat{a}_k+e^{i(\omega_k \tau -kn)}\hat{a}^{\dag}_k].   
\label{jontau}
\end{equation}
Here, $\hat{a}_k$ and $\hat{a}_{k}^{\dag}$ are Bose operators.
Using these expressions, we calculate the current-current correlator
\begin{equation} 
\Pi (k,\omega)=-i \int_0^\infty \, d \tau  {\rm e}^{i\omega \tau} 
<[\hat{\j}^{\dag}_k (\tau), \hat{\j}_k (0) ]>  
\label{curcur}
\end{equation}
and the uniform conductivity
\begin{equation}
\sigma (\omega)=-\frac{1}{\omega} \lim_{k \to 0} {\rm Im} 
\Pi (k,\omega)=2\pi e^2t\delta (\omega).
\label{sigma}
\end{equation}
The phase correlator exhibits quasi-long range order,
\begin{equation}
<\exp  i ( \hat{\varphi}_n - \hat{\varphi}_m )> 
\propto |n - m|^{- \alpha},
\label{corr}
\end{equation} 
with $\alpha = \sqrt{J/8 \pi^2 t}$. 
Hence, in the limit $t \gg J$ the average dimension of the 
K/AK pair diverges, $\xi\to \infty$, providing the
conducting ground state. Now, the elementary 
excitations are phonon-like excitations of the phase. 
This transformation is similar to what occurs in the JJ array: gapped 
charge excitations in the insulating state transform into gapless 
phase excitations in the superconducting state.  

Next, we consider the quantum dynamics of the stripe at arbitrary $t/J$.
The calculation of the complete energy spectrum corresponding to the 
string Hamiltonian (\ref{Ham2}) is a difficult task.  However, in the 
long wave-length limit $k\rightarrow 0$ the physics of the stripe can 
be described by a continuous sine-Gordon (SG) model (see Fig.1b) with 
Hamiltonian 
\begin{equation}
\hat{H}_{x}=
\int dn\left[ t\hat{p}_n^2+\frac J2\left( \frac{\partial \hat{x}_n%
}{\partial n}\right) ^2-\eta \cos (2\pi \hat{x}_n)\right] .  \label{HSG}
\end{equation}
Here, the restriction to the integer values of the coordinate $\hat{x}_n$ is
provided by the potential term $-\eta \cos (2\pi \hat{x}_n)$. 

It can be shown \cite{Itzy} that the partition function of a 2D XY model is
the imaginary time version of the action corresponding to the real time
Lagrangean 
\begin{equation}
{\cal L} = \frac{1}{2} \partial_\mu \varphi \partial^\mu \varphi - 
\frac{m^4}{\lambda} \left[1 - \cos \left( \frac{\sqrt{\lambda}}{m} \varphi
\right)\right]
\label{sineL}
\end{equation}
where we have rescaled fields and coordinates as $ x_n \to \beta \varphi /
2 \pi$ and $ (n,\tau) \to (n,\tau /c)$. Here, $\beta^2 = 
\lambda / m^2 = (2 \pi)^2 \sqrt{2 t / J}$. In Eq.\ (\ref{sineL}) $m$
is the mass of the elementary boson of the theory and $\lambda$ its coupling
constant. 

Although the equivalence of SG model to our starting 
Hamiltonian (\ref{Ham1}) is strictly correct only near criticality,
both models are dominated by K/AK excitations so that also
away from criticality, the two models should have very similar
properties.  
The SG model is further clearly a natural choice to describe
an elastic string in a periodic potential and our derivation
of the SG model from the lattice model (\ref{Ham1}) provides
us with a relation of the phenomenological parameters of the SG model
to the more microscopic parameters of the lattice Hamiltonian. 

The excitations of the SG theory are known exactly, and consist
of fermionic soliton-like excitations and bosonic bound states.
The quantization about the so-called  ``breather'' or ``doublet'' solution 
\cite{Raja} leads to a set of discrete states whose
energies are the doublet masses
\begin{equation}
M_N = 2 M_s \sin \left(\frac{N \gamma}{16} \right); \qquad
\gamma =  \frac{\beta^2}{1 - \beta^2 /8 \pi}
\label{gamma}
\end{equation}
is the renormalized coupling constant of the SG model, $M_s$ is the
soliton mass and $N = 1,2,... 8\pi / \gamma$ . In the weak coupling regime 
$\beta^2 < 4 \pi$, $M_s\simeq 8 m / \gamma$. 

These results suggest that we could regard the doublet as a bound state of
the quantum soliton-antisoliton pair. This is valid once $\gamma < 8 \pi$
$(\beta^2 < 4 \pi)$, otherwise, no bound state would survive in Eq.\ 
(\ref{gamma}). The present interpretation can be further exploited if
we use the equivalence of (\ref{sineL}) with the massive Thirring model (MTM) 
\cite{Cole} 
\begin{equation}
{\cal L}_{MT} = i \bar\Psi \gamma_\mu \partial^\mu \Psi - m_F \bar\Psi \Psi
- \frac{g}{2} (\bar\Psi \gamma^\mu \Psi) (\bar\Psi \gamma_\mu \Psi)
\end{equation}
where $\gamma_0 = \sigma_x$, $\gamma_1 = -i \sigma_y$, and $\Psi$ is a 
2-component (right and left movers) fermionic field. The constants $m_F$
and $g$ are, respectively, the mass of the fermions and the coupling constant
for their self-interaction. This equivalence allows us to identify the soliton
of the SG model with the fermion of the MTM and thereby Eq.\ 
(\ref{gamma}) can also be regarded as a set of bound states of 
fermions-antifermions of the latter.

The relationship between the coupling constants of the two models is 
\cite{Raja} $4 \pi /\beta^2 = 1 + g / \pi$
which clearly shows us that when $\beta^2 < 4 \pi$, $g > 0$. This implies that
particles and anti-particles should attract one another, in agreement with
our previous interpretation. For $\beta^2 \to 4 \pi$ one has $g \to 0$ and,
therefore, fermions and anti-fermions are about to decouple. The last
remaining bound state has mass $M_1 = 2 m / \pi$, which turns out to be twice
the solitonic mass $M_s = 8 m / \gamma$ for $\gamma = 8 \pi$. At 
$\beta^2 = 4 \pi$ fermions and antifermions are no longer bound and can
freely move along the line. 

The region where $4 \pi < \beta^2 < 8 \pi$ [see Eq.\ (\ref{gamma})] means
the very strong coupling regime of the SG theory. The coupling constant $g$
is negative, which means that particles and anti-particles should now repel
each other. The spectrum of excitations still presents a gap \cite{Emery}
that vanishes as we approach $\beta^2 = 8 \pi$ $(\gamma \to \infty)$ and,
beyond this point, the quantum mechanical SG potential becomes unbounded
below \cite{Cole}.

Actually, it has been shown by many authors \cite{KT,JeanJustin} that the 
system undergoes
a Kosterlitz-Thouless phase transition at $\beta^2 = 8 \pi$ or $g = -\pi / 2$.
Close to this transition, the SG model can be obtained from a model consisting
of two different kinds of relativistic massless fermions which is nothing but
a spinful Luttinger model (LM) \cite{JeanJustin}.
Actually, the fermions to which we are referring above can be thought of as
spin excitations of the LM, once backscattering processes are 
considered. When the latter becomes relevant, the corresponding 
excitation has the SG model as a fixed point. 

\begin{figure}[t]
\unitlength1cm
\vspace{0.1cm}
\begin{picture}(8,6)
\epsfxsize=8cm
\put(0.3,0.5){\epsfbox{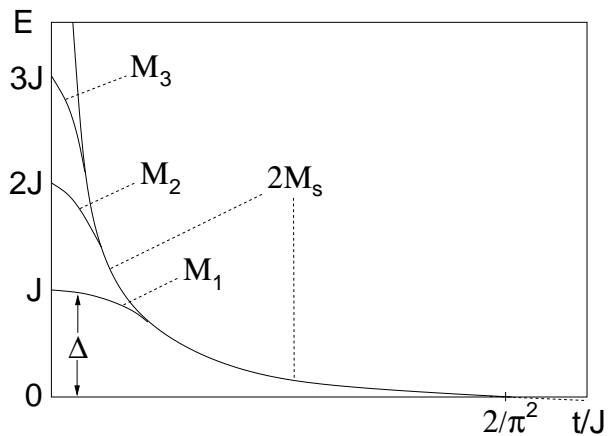}}
\end{picture}
\caption[]{\label{fig2} Infrared energy spectrum of the quantum string and the 
effective masses of its doublet excitations. Every line shows the lower
infrared boundary of continuum. The insulating gap $\Delta$ turns to zero at 
$t/J=2/\pi^2$. }
\end{figure}

In the string language (see Fig. 2), we find that for $t \ll J$, the 
excitation spectrum
is basically given by $M_N \approx N m$, which means $N$ elementary bosons
of the SG theory. It also allows us to identify $m = J$ (the lowest elementary
excitation for $t = 0$) and consequently $\lambda = (2 \pi)^2 \sqrt{2 t J^3}$.
As $t$ is increased these turn into $N$ particle bound states
which are just the excited states of the K/AK pairs. We can imagine the string
being pinned by the lattice and at least an energy $M_1$ would be necessary
to create a bound K/AK pair (the elementary boson of the SG theory). At 
$t/J =  1/2 \pi^2$ $(\beta^2 = 4 \pi)$ the pair K/AK becomes free.
When $t/J > 1 /2\pi^2$ there still
exists a gap for the formation of the pair but this bosonic gap  
vanishes as one approaches the
critical value $(t / J)_c = 2 / \pi^2$. Beyond this point the SG potential is
irrelevant, the string is no longer pinned and can freely move over the
antiferromagnetic plane. Hence, it exhibits a gaussian dynamics,
with associated logarithmic wandering. The spontaneous symmetry breaking
of the discrete system is removed and  the string becomes invariant
with respect to arbitrary transversal translations. In principle, a 
transversal sliding mode exists. The invariance of the state
to transversal translation gives rise to acoustic excitations
of the form $\omega=c^* k$ with a renormalized velocity $c^*<c$.
 
The behaviour of the correlator (\ref{corr}) is in complete 
agreement with the interpretation that above $(t / J)_c$ the spectrum
of the quantum string should be the same as for the LM we mentioned above. 
As is well known, all the correlation functions of a LM should present
algebraic decay \cite{Emery}. 
Whereas for the equivalent model of a JJ chain it really means an 
insulator-superconductor transition, here it only reflects the depinning of 
the string or, in other words, a insulator/metal transition. It would only 
require a vanishingly small electric field perpendicular to the string to 
depin it. This fact is also 
reflected in our expression (\ref{sigma}) for the perfect conductivity of the
system.

Based on the Josephson chain, as well as on the sine-Gordon results, it 
follows that at $(t/J)_c = 2 / \pi^2$
the quantum string undergoes a KT-transition. This transition has
been previously predicted \cite{Eske,Hass}, and treated as
roughening of the string.  
Besides, Vierti{\"o} and Rice \cite{Vier} have 
calculated the energy for creating a pair K/AK and have shown that for large 
$t/J$ values this energy becomes negative, leading to a proliferation of K/AK 
pairs. 
Here, we have shown that at the transition point the gap 
$\Delta$ vanishes and the bosonic excitation disappears. 
Notice that our results are based on the single stripe picture and we do not
necessarily expect them to remain valid at higher doping concentrations. 

We want to emphasize that at 
finite temperatures $(T \ne 0)$, thermal fluctuations will ``spoil'' the
quasi-long range ordered phase. 
In this case, the Euclidean action (\ref{Ecaction}) describes a XY 
model on a 2D lattice, which is finite in the $\tau$-direction, with length 
 $L = 2 \pi /T$. Then, the KT-transition disappears and the long-range
phase correlations are suppressed. 

Finally, we can summarize our results: at $t = 0$, the GS of the string is the
kink-vacuum. At $0<t/J < 2 /\pi^2$, the energy spectrum is gapped and the 
system is insulating.  At $t/J > 2/ \pi^2$ 
there is no gap anymore, the phase is quasi long-range ordered, the GS is
the one of the LM 
and the system is a perfect conductor. Thus, our results at the critical
region agree with the ones obtained from the mapping onto the 
Josephson chain, with the advantage that they clarify the physical meaning of
the insulating and superconducting states for the quantum string. 
Besides, we are proposing a phenomenological model which provides us with the
spectrum of a quantum string for {\it any value} of $t / J$.

We are indebt with H.\ Schmidt, T.\ M.\ Rice, A.\ H.\ 
Castro Neto and D. Baeriswyl for fruitful
discussions. This work has been supported by the DAAD-CAPES project number 
415-probral/sch\"u (Germany) and 053/97 (Brazil). N.\ H.\ acknowledges 
financial support from the Gottlieb Daimler- und Karl Benz-Stiftung and the 
Graduiertenkolleg ``Physik nanostrukturierter Festk\"orper'', 
Universit\"at Hamburg.  Y.\ D.\ acknowledges financial support from the Otto
Benecke-Stiftung. A.\ O.\ C.\ was also partly supported by CNPq (Conselho 
Nacional de Desenvolvimento Cient\'{\i}fico e Tecnol\'{o}gico / Brasil).

\end{document}